\documentclass[apjl,iop]{emulateapj}
\slugcomment{The Astrophysical Journal Letters, 771: L44 (7pp), 2013 July 10} 
\usepackage[table]{xcolor}
\usepackage{graphicx}
\usepackage[normalem]{ulem}
\usepackage{rotating}

\shortauthors{Pasham, Strohmayer \& Mushotzky}
\begin{document}
\title{Discovery of a 7 {\lowercase{m}}H\lowercase{z} X-ray Quasi-periodic Oscillation from \\ the most massive
stellar-mass black hole IC 10 X-1}
\author{Dheeraj R. Pasham\altaffilmark{1,2}, Tod E. Storhmayer\altaffilmark{2}, Richard F. Mushotzky\altaffilmark{1}}
\affil{$^{1}$Astronomy Department, University of Maryland, College Park, MD 20742; email: dheeraj@astro.umd.edu; richard@astro.umd.edu \\ 
$^{2}$Astrophysics Science Division, NASA's GSFC, Greenbelt, MD 20771; email: tod.strohmayer@nasa.gov \\
{\it Received 2013 May 31; Accepted 2013 June 14; Published 2013 June 27}
}

\begin{abstract}
We report the discovery with {\it XMM-Newton} of an $\approx$ 7 mHz
X-ray (0.3-10.0 keV) quasi-periodic oscillation (QPO) from the
eclipsing, high-inclination black hole binary IC 10 X-1. The QPO is 
significant at $>$ 4.33$\sigma$ confidence level and has a fractional 
amplitude (\% rms) and a quality factor, $Q\equiv \nu /
\Delta\nu$, of $\approx 11$ and $4$, respectively. The
overall X-ray (0.3-10.0 keV) power spectrum in the frequency range
0.0001 - 0.1 Hz can be described by a power-law with an index of
$\approx -2$, and a QPO at 7 mHz. At frequencies $\ga$ 0.02 Hz there
is no evidence for significant variability. The fractional amplitude
(rms) of the QPO is roughly energy-independent in the energy range of 
0.3-1.5 keV. Above 1.5 keV the low signal to noise ratio of the data
does not allow us to detect the QPO.  By directly comparing these
properties with the wide range of QPOs currently known from accreting
black hole and neutron stars, we suggest that the 7 mHz QPO of IC 10 X-1 may be linked
to one of the following three categories of QPOs: (1) the
``heartbeat'' mHz QPOs of the black hole sources GRS 1915+105 and IGR
J17091$-$3624, or (2) the 0.6-2.4 Hz ``dipper QPOs'' of
high-inclination neutron star systems, or (3) the mHz QPOs of Cygnus
X-3.
\end{abstract}



\newpage
\section{Introduction}
The X-ray light curves of numerous accreting neutron star and
stellar-mass black holes (StMBHs) show evidence for the presence of
quasi-periodic oscillations (QPOs), which appear as finite-width peaks
in their power density spectra (PDS) (see van der Klis 2006 and
McClintock \& Remillard 2006 for reviews of neutron star and StMBH
QPOs). While it is known that QPOs occur with a wide
range of centroid frequencies--a few mHz to above a kHz in neutron
stars and a few mHz to a few hundred Hz in the case of StMBHs--the
exact nature of the physical processes producing such oscillations is
still a mystery.

Based on the observed properties, that is, their centroid frequencies,
widths, amplitudes, and overall nature of their power spectra, etc.,
QPOs have been categorized into different groups. In neutron star
binaries the QPO phenomenon constitutes the {\it kilohertz} QPOs
(centroid frequencies in the range of 300-1200 Hz: see the review by
van der Klis 2000) seen from over two dozen sources (e.g., M{\'e}ndez
et al. 2001; Barret et al. 2008 and references therein), the {\it
hectohertz} QPOs ($\sim$100-300 Hz: e.g., van Straaten et al. 2003;
Altamirano et al. 2008a) seen predominantly in a special class (atoll)
of neutron star binaries, the low-frequency QPOs (0.01-50 Hz: e.g.,
van Straaten et al. 2003), the 1 Hz QPOs observed in two accreting
millisecond X-ray pulsars (e.g., Wijnands 2004), the $\approx 0.6-2.4$
Hz QPOs observed only from dipping (high-inclination) neutron star
binaries (e.g., Homan et al. 1999; Jonker et al. 1999, 2000) and the very low-frequency 7-15 mHz QPOs
observed from at least three systems (e.g., Revnivtsev et al. 2001, Altamirano et al. 2008b).

Similarly, black holes also show a variety of QPOs (McClintock \&
Remillard 2006). They can be broadly classified into two categories:
(1) high-frequency QPOs (HFQPOs) with centroid frequencies in the
range of a few$\times$(10-100) Hz (e.g., Miller et al. 2001; Strohmayer 2001; Remillard et al. 2006; Belloni \& Altamirano 2013) and (2)
low-frequency QPOs (LFQPOs) that occur in the range of 0.1-15 Hz
(e.g., Casella et al. 2005). Based on their broadband properties,
viz., shape, fractional amplitude of the PDS and the QPOs, etc., the
LFQPOs have been further sub-divided into type-A, B and C (e.g.,
 Homan et al. 2001; Remillard et al. 2002). In addition to the HFQPOs
and the LFQPOs of StMBHs, two black hole sources, GRS 1915+105 and IGR
J17091$-$3624, show so-called ``heartbeat'' QPOs which occur in the
mHz frequency regime (e.g., Belloni et al. 2000; Altamirano et
al. 2011). Furthermore, some ultraluminous X-ray sources (ULXs) show a
few$\times$10 mHz QPOs (e.g., Dewangan et al. 2006; Pasham \&
Strohmayer 2013). More recently, an 11 mHz X-ray QPO and the
recurrence of a few$\times$mHz QPOs were detected from the black hole
candidates H1743$-$322 and Cygnus X-3, respectively (Koljonen et al. 2011; Altamirano \&
Strohmayer 2012).

IC 10 X-1 is an eclipsing, Wolf-Rayet binary containing the most
massive StMBH known with an estimated black hole mass of 23-34 M$_{\odot}$
(Prestwich et al. 2007; Silverman \& Filippenko 2008). The presence of
an eclipse suggests that the system is highly inclined, i.e., close to
edge-on. This source was observed previously with {\it XMM-Newton}
(ID: 0152260101) for a duration of roughly 45 ks. After accounting for
background flaring only a mere 15 ks of useful data was available, 
analysis of which showed some evidence--although
at modest statistical significance--for the presence of a QPO at
$\approx$ 7 mHz. Motivated by this, and to carry out eclipse mapping,
a long {\it XMM-Newton} observation was proposed to confirm the presence of
this mHz QPO (ID: 0693390101; PI: Strohmayer). Here we present results
from our timing analysis of this new data set and confirm the presence
of the QPO at 7 mHz.


\begin{figure}[ht!]
\includegraphics[width=3.5in, height=2.85in, angle=0]{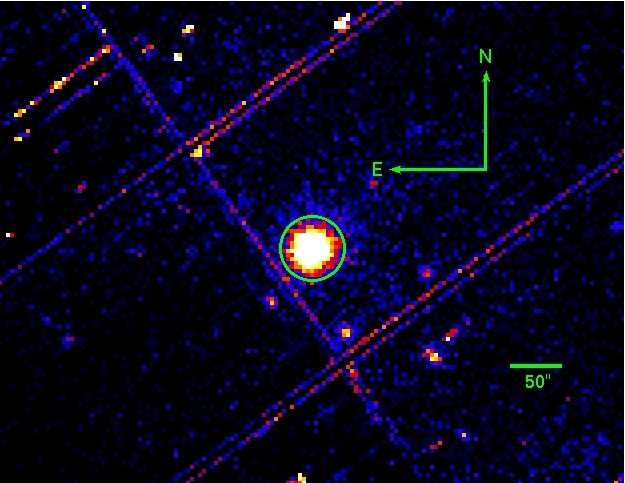}

\caption{EPIC-pn X-ray (0.3-10.0 keV) image of IC 10 X-1. Clearly there is only one point source, IC 10 X-1, and no obvious evidence for source contamination. The extraction region of radius 33'' is indicated by a green circle. }
\label{fig:figure2}
\end{figure}


\section{ {\it XMM-Newton} observations}
Beginning 2012 August 18 at 22:05:46 (UTC) {\it XMM-Newton} observed
IC 10 X-1 for roughly 135 ks, a duration approximately equal to its
orbital period (34.93 hrs: Prestwich et al. 2007; Silverman \& Filippenko 2008). For our study we only used the EPIC data
(both pn and MOS). We used the latest standard analysis system 
version 13.0.0 for extracting the images and the event lists. The
source was easily identifiable and there were no source confusion
problems (see Figure 1). The source events were extracted from a circular region of
radius 33'' centered around the source and the background events were
extracted from a nearby circular region of the same size. The
observation was affected by flaring only briefly at the very beginning
and the end of the pointing. These epochs were removed from our
analysis.

The combined pn and MOS 0.3-10.0 keV light curve of IC 10 X-1 is shown
in Figure 2 ({\it black}) along with the background ({\it red}). Also
overlaid are the good time intervals (GTIs) during which a given EPIC
instrument (pn/MOS1/MOS2) was continuously active for more than 5
ks. For a given instrument the horizontal line, which is offset to an
arbitrary value, indicates the active time, while a vertical line
marks the beginning or the end of a continuous GTI. It is clear that
EPIC-pn has three GTIs of duration roughly 23 ks, 75 ks and 27 ks,
while MOS1 has two GTIs of length 30 ks and 99 ks and MOS2 has one
long GTI of 130 ks.

\section{Results}
It is clear even by eye that the source varies significantly. Since
the pn detector offers the highest effective area among the three EPIC
instruments, we started our analysis with its longest available GTI of
$\approx$ 75 ks (between hour 8.5 to 29.5 in Figure 2). Using all the
0.3-10.0 keV photons we constructed a Leahy-normalized PDS where the
Poisson noise equals 2 (Leahy et al. 1983). This is shown in the top left
panel of Figure 3 ({\it histogram}). It is evident that the overall
power spectrum can be described by a simple power-law noise at the
lowest frequencies with a QPO-like feature around 7 mHz and
essentially Poisson noise at frequencies above $\ga$ 0.02 Hz. In order
to test the significance of the QPO we followed a rigorous Monte Carlo
approach described below.

First, we fit the continuum of the PDS using a model consisting of a
power-law plus a constant. While modeling the continuum we used the
frequency range 0.0001 (the lowest that can be probed) - 0.1 Hz and
excluded the region containing the apparent QPO feature, i.e., 5-9
mHz. The best-fit continuum model parameters are shown in the first
column of Table 1. Thereafter, following the prescription described by
Timmer \& Koenig (1995), we simulated a large number of light curves
(and their corresponding PDS) that have the same shape, i.e., same
parameters, as in the 1$^{st}$ column of Table 1, and the same
frequency resolution as the spectrum used for obtaining the best-fit
continuum parameters. A sample PDS simulated with the above technique
({\it red}) along with the real PDS ({\it black}) is shown in the top 
right panel of Figure 3. We simulated 370 such power spectra and found
the maximum value in each frequency bin. This gave us the 99.73\%
(3$\sigma$) significance within that particular bin. A similar
estimate for each frequency bin gave us the complete confidence
curve. Similarly we simulated 10000 PDS and estimated the 99.99\%
(3.9$\sigma$) confidence curve. It should be noted that the confidence
curves are sensitive to the chosen values of the continuum model
parameters. Given the error on each of the individual model
parameters, i.e., best-fit power-law normalization and the power-law
index, we estimated the 99.73\% and the 99.99\% curves for various
combinations of the power-law normalization and the index within the
error bars quoted in column 1 of Table 1. To be conservative we picked
the maximum of these curves. These confidence levels are overlaid in
the figure. It is clear that the QPO feature is significant at the
99.99\% level.


%
%
\begin{figure*}
  \begin{center}
 \includegraphics[width=17.5cm]{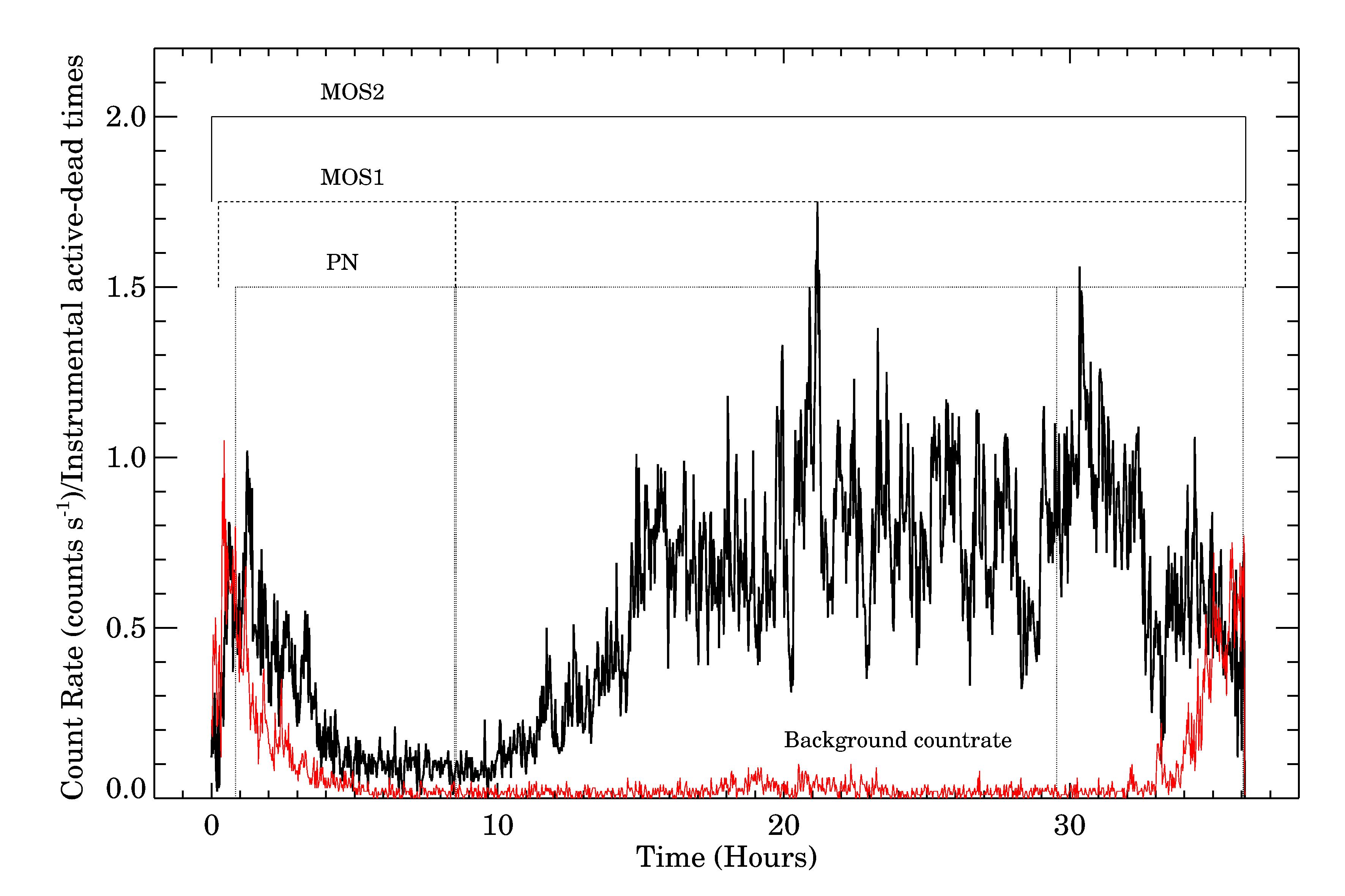}    
  \caption[Figure2]{The combined, background-subtracted EPIC X-ray (0.3-10.0 keV) light curve of IC 10 X-1 ({\it black}) along with the background light curve ({\it red}). Time zero corresponds to 4.6171485$\times$10$^{8}$ secs since 50814.0 (Modified Julian Date). Both the light curves were binned to 100 seconds. The start and end times of all the GTIs greater than 5 ks are indicated by vertical lines (see text).}
     \label{fig:hubble_dia}
  \end{center}
\end{figure*}


To further confirm the presence of this QPO feature we extracted
another PDS using the combined MOS data. For this purpose we used the
longer GTI of roughly 95 ks (from hour 8.5 to hour 34.5 in Figure
2). The 0.3-10.0 keV combined MOS PDS is shown in the bottom left panel of
Figure 3. The QPO feature is again evident at 7 mHz. To quantify the
variability we first modeled the PDS with a power-law plus a
constant. This gave a $\chi^2$ of 245 with 192 degrees of freedom
(dof). We then added a Lorentzian component to model the QPO feature
at 7 mHz. This improved the $\chi^2$ by 28 with an addition of three
parameters, i.e., a $\chi^2$ of 217 with 189 dof (see the second
column of Table 1). This decrease in $\chi^2$ serves as a further
indicator of the significance of the QPO component. Using the F-test this corresponds to a single-trial significance of
4$\times$10$^{-5}$. Note that we essentially searched in a known narrow 
frequency range of $\approx$ 5-10 mHz (from the prior {\it XMM-Newton} observation), 
and thus the effective number of trials is $\sim$ 1.

It is clear that the QPO is present in two independent detectors at
$>$ 3.5$\sigma$ confidence level in each case. The chance probability
of two independent 3$\sigma$ detections alone is 4.33$\sigma$. Given
the 3.5$\sigma$ detections in two separate measurements, we conclude
that the observed 7 mHz QPO of IC 10 X-1 is statistically highly
significant.


\begin{table*}
  \begin{flushleft}
\centering
  \caption{{\normalsize Summary of power spectral modeling.}}\label{Table1} 
\vspace{.25cm}
{\footnotesize
    \begin{tabular}[t]{lccccccc}
	\\
    \hline\hline \\
   Parameter				& EPIC-pn 	   	& Combined MOS   & Combined MOS    & Combined MOS	 & Combined MOS    &   Combined MOS \\
					& (Continuum)	   	& (0.3-10.0 keV) &  (0.6-10.0 keV) &(0.9-10.0 keV)    &(1.2-10.0 keV)   & (1.5-10.0 keV)  \\
\\
    \hline \\
\\
N($\times$10$^{-7}$)\tablenotemark{a}& $143 \pm 125$   	& $1.6 \pm 1.7$  & $1.8 \pm 1.9$   & $1.3 \pm 1.5$   & $3.2 \pm 3.5$   & $1.2 \pm 1.5$  \\ 
\\
$\Gamma$\tablenotemark{a}		& $1.9 \pm 0.1$  	& $2.4 \pm 0.2$  & $2.4 \pm 0.2$   & $2.5 \pm 0.2$   & $2.3 \pm 0.2$   & $2.4 \pm 0.2$ \\   
\\
C\tablenotemark{a}			& $1.9 \pm 0.1$  	& $1.9 \pm 0.1$  & $1.9 \pm 0.1$   & $1.9 \pm 0.1$   & $1.9 \pm 0.1$   & $1.9 \pm 0.1$   \\ 
\\
N$_{QPO}$\tablenotemark{b}		& --\tablenotemark{e} & $1.5 \pm 0.4$  & $1.5 \pm 0.4$   & $1.4 \pm 0.4$   & $1.4 \pm 0.4$   & $1.0 \pm 0.4$   \\
\\
$\nu_{0}$(mHz)\tablenotemark{b} 	& --\tablenotemark{e}	& $6.3 \pm 0.2$  & $6.3 \pm 0.2$   & $6.2 \pm 0.2$   & $6.3 \pm 0.2$   & $6.3 \pm 0.2$  \\  
\\
$\Delta\nu$(mHz)\tablenotemark{b}	& --\tablenotemark{e}	& $1.5 \pm 0.5$  & $1.5 \pm 0.5$   & $1.7 \pm 0.6$   & $1.6 \pm 0.6$   & $1.5 \pm 0.7$  \\
\\
Q\tablenotemark{c} 			& --\tablenotemark{e}	& 4.2		 & 4.2              & 3.7	      & 3.9		 & 4.2            \\
\\
RMS$_{QPO}$\tablenotemark{d}		& --\tablenotemark{e}	& $11.1 \pm 2.5$ & $11.2 \pm 2.5$  & $11.9 \pm 2.7$  & $12.6 \pm 2.9$  & $12.0 \pm 3.5$  \\
\\
\hline
\\
$\chi^2$/dof            		& 354/284               & 216/189        & 219/189         &  215/189         & 243/189          & 220/189          \\
\\
    \hline\hline
    \end{tabular}
\\
}
  \end{flushleft}

\vspace{-0.25cm}
{\scriptsize 
\tablenotemark{a}{We fit the continuum with a power-law model described as follows: \\
\begin{center}
\begin{math} Continuum = N\nu^{-\Gamma}+C\end{math}
\end{center}
where, $\Gamma$ is the power-law index of the continuum.
}\\
\tablenotemark{b}{We modeled the QPOs with a Lorentzian. The functional form is as follows: 
\begin{center}
\begin{math} QPO = \frac {N_{QPO}} {1 + \left(\frac {2(\nu - \nu_{0})} {\Delta\nu}\right)^{2} } \end{math}
\end{center}
where, $\nu_{0}$ is the centroid frequency and $\Delta$$\nu$ is the full-width-half-maximum (FWHM) of the QPO feature.\\
\tablenotemark{c}{The quality factor of the QPO defined as $\nu_{0}$/$\Delta$$\nu$.}\\
\tablenotemark{d}{The fractional RMS amplitude of the QPO (see text).}\\
\tablenotemark{e}{In this case we only modeled the continuum.}\\
}
}
\end{table*}


In addition, we studied the energy dependence of the fractional RMS
amplitude of the QPO. For this purpose, we extracted the PDS of the
source in seven energy bands. Owing to the low count rate we fixed the
upper bound of the bandpass at 10 keV and varied the lower limit from
0.3 to 1.5 keV and constructed a PDS in each case using the combined
MOS data. Each of these PDS were then modeled with a power-law plus
constant for the continuum and a Lorentzian for the QPO (best-fit
model parameters shown in Table 1). The fractional RMS amplitude of
the QPO is:
\[ 
RMS ~ amplitude ~(\%) = 100\left( \sqrt{\frac{\pi N W}{2C}} \right)
\]
where $N$ and $W$ are the normalization and the width of the QPO
(Lorentzian), respectively, while $C$ is the mean count rate of the
source. The dependence of the RMS amplitude of the QPO as a function
of the lower limit of the band pass is shown in the bottom right panel of
Figure 3. There is a very weak dependence of the QPO's amplitude on
the energy from 0.3-1.5 keV. Four of the five PDS used for this
analysis are shown in Figure 4. At energies greater than
1.5 keV the low signal to noise ratio of the data does not allow us to
detect the QPO.

\subsection{Search for a power spectral break}
Numerous X-ray binaries and also active galactic nuclei show evidence
for the presence of a break in their PDS (e.g., McHardy et al. 2006;
Markowitz \& Edelson 2004). We searched for a spectral break in IC 10
X-1 using the data from the longest GTI outside the eclipse, i.e.,
hour 15.5 to 33.5 in Figure 2. Note that the presence of an eclipse in
the data adds red noise to the power spectrum that is not intrinsic to
the source variability. We constructed the combined MOS 0.3-10.0 keV
PDS and did not find any obvious evidence for a PDS break down to
frequencies as low as 0.0001 Hz. Note that a single PDS is noisy, with
error in a particular bin equal to the value of that bin (van der Klis
1989). Therefore, averaging (say, by combining neighboring bins) is
necessary to reduce the noise in the PDS. Hence, even though the
lowest sampled frequency is $\approx$ 10$^{-5}$ Hz (1/total length)
averaging reduces the lowest effective frequency to roughly 0.0001 Hz
in this case. It remains possible that a break may exist at $\la$
0.0001 Hz. Moreover, we modeled this PDS with a power-law and a
Lorentzian. We find that the best-fit continuum can be described by a
power-law of index $\approx -2$.


%
%
\begin{figure*}
  \begin{center}
	\includegraphics[width=16cm]{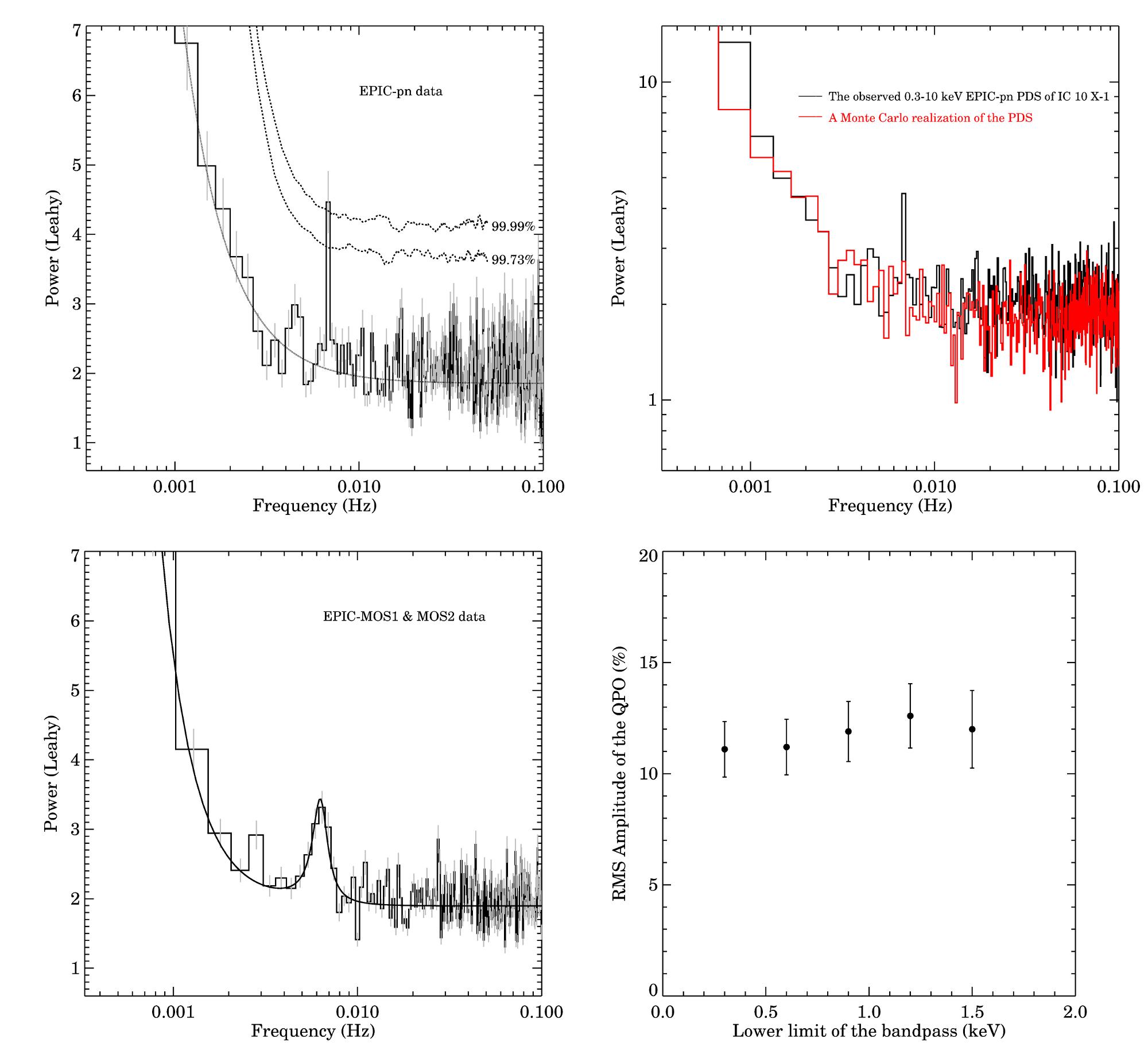}    
       \caption[Figure3]{Top Left panel: The EPIC-pn PDS of IC 10 X-1 using the longest GTI of 75 ks ({\it black histogram}) along with the best-fit power-law model for the continuum ({\it solid}). The 99.73\% and the 99.99\% Monte Carlo simulated confidence contours are also shown ({\it dashed}). The QPO at 7 mHz is evident. Top Right panel: The observed EPIC-pn PDS of IC 10 X-1 ({\it black}: same as the figure on the left) and a sample Monte Carlo simulated PDS ({\it red}). Bottom Left panel: The combined 0.3-10.0 keV EPIC-MOS PDS of IC 10 X-1 using the longest common GTI of 95 ks ({\it histogram}). The best-fit model is also shown ({\it solid}). Again the feature at 7 mHz is evident. Bottom Right panel: The fractional RMS amplitude of the QPO versus the lower bound of the band pass used for constructing the PDS. The upper limit was fixed at 10 keV (see text).}
     \label{fig:figure3}
  \end{center}
\end{figure*}


\section{Discussion}
The frequency of the HFQPOs of StMBHs ($\sim$ a few 100 Hz) and the {\it hectohertz}
QPOs of neutron stars ($\sim$ 100-300 Hz) are roughly constant in
frequency for a given source (van der Klis 2006). They are thought to
have a common origin (e.g., Abramowicz et al. 2003) and it has been
proposed that the QPO frequency may scale inversely with the mass of
the compact object (e.g., see Figure 4.17 of McClintock \& Remillard
2006). With a mass of 23-34 M$_{\odot}$ IC 10 X-1's HFQPOs, if any,
are expected to occur in the range of a few 10s of Hz. Clearly the 7
mHz QPO of IC 10 X-1 is orders of magnitude slower than this and is
very likely not a HFQPO phenomenon.

The typical values of the centroid frequency, RMS amplitude and the
quality factor ($Q =$ centroid-frequency/QPO-width) of type-A LFQPOs
are $\sim 8$ Hz, $\la 3$ and $\la 3$, respectively (e.g., Casella et
al. 2005). The respective values for type-B LFQPOs are $\sim 5-6$ Hz,
$\sim 2-4$ and $\ga 6$. Lastly, type-C QPOs occur in a wider range of
frequencies -- 0.1-15 Hz -- and have RMS amplitudes of 3-20 with $Q$
factors of $\sim 7-12$. The overall nature of the PDS accompanying
type-A, B, and C QPOs can be described as weak red noise, weak red
noise and strong flat-topped noise, respectively. Although the
continuum of the PDS of IC 10 X-1 is similar to that accompanying
type-A or B QPOs, its QPO frequency, RMS amplitude and $Q$ value are
quite different (compare values in Table 1 with Table 1 of Casella et
al. 2005). On the other hand, the centroid frequency of IC 10 X-1's
QPO is much lower compared to a typical type-C QPO.


%
%
\begin{figure*}
  \begin{center}
	\includegraphics[width=16cm]{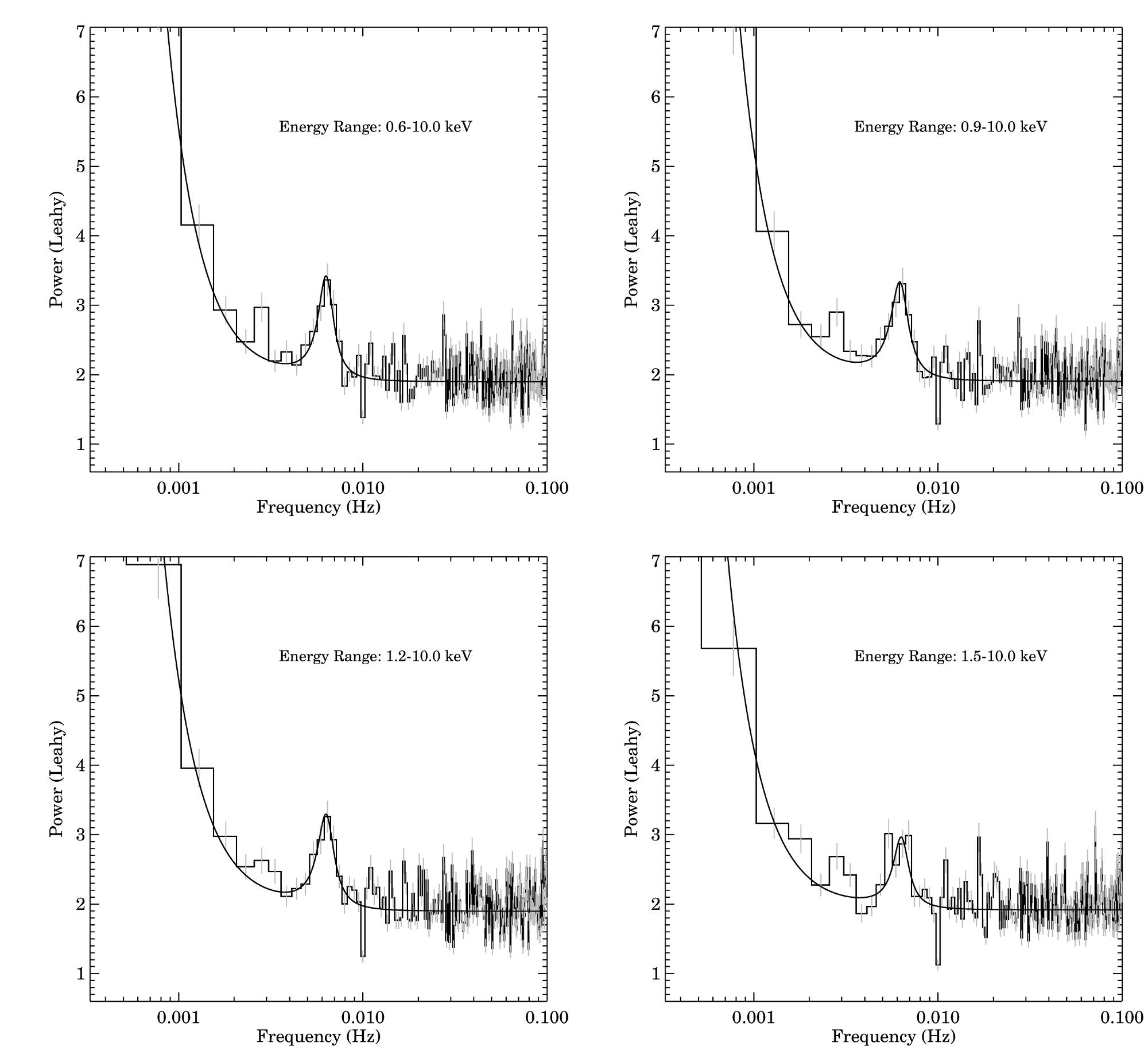}    
       \caption[Figure4]{Combined MOS PDS of IC 10 X-1 in
various energy bands. The energy band used for the PDS is indicated at
the top-right of each panel. The best-fit model (see Table 1) is also
overlaid ({\it solid}) in each case.}
     \label{fig:figure4}
  \end{center}
\end{figure*}


The mHz QPOs (frequency range of $\sim$ 10-200 mHz) of ULXs have been
argued to be the analogs of the type-C LFQPOs of StMBHs but occurring
at a lower frequency (a few 10s of mHz compared to the few Hz of
StMBHs) due to the presence of intermediate-mass black holes (mass of
a few$\times$(100-1000) M$_{\odot}$) within these systems. While the
centroid frequency, the RMS amplitude and the $Q$ value of the 7 mHz
QPO of IC 10 X-1 are comparable to the mHz QPOs of ULXs (e.g., Dheeraj
\& Strohmayer 2012), there are two aspects that are dissimilar. (1) We
do not detect a break in the PDS of IC 10 X-1 whereas breaks have been
seen in all the ULXs (e.g., Dewangan et al. 2006). It is known that the break
frequency scales with the QPO frequency as $\nu_{break}$ $\sim$
$\nu_{QPO}$/9 (Wijnands \& van der Klis 1999). If that were the case
for IC 10 X-1, the expected break is at $\sim$ 0.7 mHz. It is thus
possible that we are unable to detect the break due to our inability 
to sample variability at very low ($\la$ 0.7 mHz) frequencies, or the effects of the
eclipse. (2) The RMS amplitude--at least in the case of the ULX NGC
5408 X-1--is known to increase with energy from 0.3-2.0 keV
(Strohmayer et al. 2007; Middleton et al. 2011). However, we do not
find evidence for such behavior in IC 10 X-1 (see the right panel of Figure 3).

Two black hole sources, GRS 1915+105 and IGR J17091$-$3624, show mHz
QPOs in the so-called ``heartbeat'' state or the $\rho$ state (e.g., Greiner et al. 1996; Morgan et al. 1997; Belloni et al. 2000;
Altamirano et al. 2011). These QPOs are thought to be the result of a
radiation pressure instability causing quasi-periodic evaporation
followed by refilling of the inner regions of the accretion disk
(Lightman \& Eardley 1974, Belloni et al. 1997; Neilsen 2011). These mHz QPOs occur at relatively high
luminosities ($\sim$ 10$^{38}$ erg s$^{-1}$) and at least in GRS
1915+105 appear to be energy-independent in the band pass from 2-30
keV (see Figure 8 of Morgan et al. 1997). Given the similar frequency,
comparable RMS amplitude (e.g., Altamirano et al. 2011), energy
independence (although here energy independence is seen over a
different X-ray band pass) and similar X-ray luminosity\footnote{The X-ray (0.3-10.0 keV) energy spectrum of IC 10 X-1 outside the eclipse can be fit with a canonical model consisting of a disk-blackbody and a power-law plus a Gaussian to model the emission feature at $\approx$ 0.9 keV. This model gives an acceptable fit with a $\chi^2$ of 165 for 125 degrees of freedom. A detailed spectral analysis is not the subject of this work. Nevertheless, assuming this simple model, the inferred 2-10 keV luminosity at a distance of 0.66 Mpc is $\sim$ 10$^{38}$ ergs s$^{-1}$.} (see also Wang et
al. 2005) it is possible that the 7 mHz QPO of IC 10 X-1 is related to
the ``heartbeat'' QPOs. Given its low count rate it is, however, not
possible to resolve IC 10 X-1's light curve to the same level as GRS
1915+105 or IGR J17091$-$3624. It will require instruments with
larger collecting area to test this hypothesis.

The 7 mHz QPO of IC 10 X-1 is likely not related to the 1 Hz QPOs of
accreting millisecond X-ray pulsars (AMXPs) or the 7-15 mHz QPOs seen
in some neutron star systems (e.g., Revnivtsev et al. 2001, Altamirano et al. 2008b). The 1 Hz QPO phenomenon in AMXPs is
thought to be due to disk instabilities within the boundary layer of
the accretion disk and neutron star magnetosphere (Patruno et
al. 2009). Moreover, the 1 Hz QPOs are seen at low luminosities ($<$
10$^{36}$ erg s$^{-1}$) and can have RMS amplitudes as large as 100\%
(Patruno et al. 2009). Given the high RMS amplitude, and that they are
likely related to the beginning of the propeller regime, they are
probably different from the QPO seen in IC 10 X-1. The 7-15 mHz
oscillations in some neutron stars are linked to marginally stable
burning of light elements on the surface of the neutron star (e.g.,
Heger et al. 2007), a process unique to neutron stars.

On the other hand, the 7 mHz QPO may be connected to the 0.6-2.4 Hz
``dipper QPOs'' of high-inclination neutron stars in the sense that IC
10 X-1 is also highly inclined (see the eclipse in Figure 1 and
Silverman \& Filippenko 2008). The so-called ``dipper QPOs'' are only
seen from X-ray dipping sources. The dipping is presumably due to
obscuration associated with the high inclination (Parmar \& White
1988). Their RMS amplitudes are $\sim$ 10\% and are energy-independent
in the range of 2-30 keV (see, for example, Figure 4 of Homan et
al. 1999). With the present {\it XMM-Newton} data a similar energy
range cannot be probed. However, we note that the QPO's RMS amplitude
is comparable to those of ``dipper QPOs'' and appears to be
independent of energy in the range from 0.3-1.5 keV. More recently
Altamirano \& Strohmayer (2012) reported the discovery of an 11 mHz
QPO from a black hole candidate H1743-322 (likely highly
inclined: Homan et al. 2005) and suggested this could be the first
detection of a ``dipper QPO'' analog in a black hole (candidate)
system. The centroid frequency of the ``dipper QPOs'' is roughly
constant over time. If the 7 mHz QPO is indeed a ``dipper QPO'' then
it's centroid frequency should also remain more or less constant. This
can be tested with multi-epoch observations of IC 10 X-1 to search for
QPO variability.

Finally, we note that the overall PDS and the QPO properties of IC 10
X-1 are also similar to that of Cygnus X-3 (van der Klis \& Jansen 1985). 
They both have the same power-law like noise at low frequencies--each 
with roughly the same slope of $-2$--with a QPO in the mHz regime and 
barely any power above 0.1 Hz (see the bottom panels of Figure 2 of 
Koljonen et al. 2011). In addition the QPOs in both cases have 
comparable frequencies, RMS amplitudes and coherences 
(van der Klis \& Jansen 1985). The mHz oscillations of Cygnus X-3 are
likely associated with major radio flaring events (Koljonen et
al. 2011; see, for example, Lozinskaya \& Moiseev (2007) and references therein for radio studies of IC 10 X-1). Simultaneous radio and X-ray observations in the future can
test whether jet ejection is related to the QPO in IC 10 X-1.

We would like to thank the referee for a timely and helpful report.


\vfill\eject

\end{document}